\documentclass{PoS}
\usepackage{graphicx}
\usepackage{times}
\usepackage{epsfig}
\newcommand{\be}{\begin{equation}}
\newcommand{\ee}{\end{equation}}
\newcommand{\beqa}{\begin{eqnarray}}
\newcommand{\eeqa}{\end{eqnarray}}
\title{Precision SM calculations and theoretical interests beyond the SM in $K_{\ell2}$ and $K_{\ell3}$ decays}

\ShortTitle{Theoretical interests in $K_{\ell2}$ and $K_{\ell3}$ decays}

\author{\speaker{Emilie Passemar}\thanks{Address after the $1^{st}$ of October: 
IFIC, Universitat de Val\`encia - CSIC, Apt. Correus 22085, E-46071 Val\`encia, Spain.}\\
Albert Einstein Center for Fundamental Physics \\
Institute for Theoretical Physics \\
Sidlerstrasse 5 \\
CH-3012 Bern \\
Switzerland 
\\
 E-mail: \email{passemar@itp.unibe.ch}}

\abstract{We present a brief overview of the recent theoretical progress to describe the $K_{\ell 2}$ and 
$K_{\ell 3}$ decays. We discuss the interesting probes of the Standard Model offered by these decays such as the extraction 
of $|V_{us}|$ and the test of the CKM unitarity, the tests of lepton universality, the quark mass ratio determination and 
the test of the electroweak couplings of the light quarks to the W-boson.}

\FullConference{2009 KAON International Conference KAON09,\\
		 June 09 - 12 2009\\
		 Tsukuba, Japan}

\begin{document}

\section{Introduction}
Despite the great success of the Standard Model (SM) in describing all the low energy measurements so far, it is 
generally believed that it only represents the low energy limit of a more fundamental theory. 
There exist two main roads to look for physics beyond the Standard Model: 
direct searches for new particles (Charged Higgs, Supersymmetric particles, Z', W'...) at high energy 
colliders and indirect searches, for instance in flavour physics, through precision experiments, 
providing sensitivity to the new degrees of freedom at high energy which are suppressed at low energy. 
For such high precision studies, we will discuss in the following two low-energy processes 
namely the $K_{\ell 2}$ and $K_{\ell 3}$ decays.  
After having reviewed the necessary theoretical tools to study them, we will present 
the stringent tests of the SM that their measurements offer. 
\section{Theoretical framework}
\subsection{$K_{\ell3}$ decays}
The $K_{\ell 3}$ decays represent the gold plated channel to extract $|V_{us}|$ 
with an impressive precision. This is done by measuring, for the four modes 
($K = K^{\pm}$, $K^0$ and $\ell = \mu$, $e$), the $K_{\ell3}$ decay rates conveniently decomposed as
\begin{eqnarray}
\label{eq:Mkl3}
\Gamma_{K_{\ell 3(\gamma)}} ={ G_F^2 M_K^5 \over 192 \pi^3} C_K^2 S_{\rm EW}\, \Big|  f_+^{K^0 \pi^-}(0) V_{us} \Big|^2\,
I_K^\ell \, \left(1 + 2 \Delta^{K \ell}_{\rm EM}
+ 2 \Delta^{K \pi}_{\rm SU(2)} \right)~,
\end{eqnarray}
where $G_F$ is the Fermi constant extracted from muon decays, 
$S_{EW}= 1+ \frac{2\alpha}{\pi} \left( 1- 4 \frac{\alpha_s}{4 \pi}\right) \rm {log} \left( \frac{M_Z}{M_\rho} \right) + 
\mathcal{O} \left( \frac{\alpha_S \alpha}{4 \pi}\right)=1.0232(3)$ \cite{Sirlin:1977sv,Marciano:1993sh} denotes the short 
distance electroweak correction, 
$C_{K}$ the Clebsh Gordan coefficient equal to $1$ ($1/\sqrt{2}$) for the neutral (charged) kaon decays,
$f_{+}^{K^0 \pi^-} (0)$ the $K^0 \to \pi^-$ vector form factor at zero 
momentum transfer, and 
$I_K^\ell$ the phase space integral which  
depends on the form factor parameters (slope, curvature...).
To compute the latter, we need information about the  $t$-dependence
of the form factors
defined by the QCD matrix elements
\begin{eqnarray}
\langle \pi(p_\pi) | \bar{s}\gamma_{\mu}u | K(p_K)\rangle =
(p_\pi+p_K)_\mu\  f^{K \pi}_+ (t) + (p_K-p_\pi)_\mu\  f_-^{K \pi } (t)~,         
\label{eq:hadronic element}
\end{eqnarray}
where $t=(p_K-p_\pi)^2= (p_\ell+p_\nu)^2$. 
The vector form factor $f^{K\pi}_+(t)$ represents the P-wave projection of the crossed channel 
matrix element $ \langle 0 |\bar{s}\gamma^{\mu}u| K \pi \rangle $, whereas the S-wave projection 
is described by the scalar form factor defined as
\begin{equation}
f^{K \pi}_0(t)= f^{K \pi}_+(t) + \frac{t}{M_K^2-M_\pi^2} f^{K \pi}_-(t)~.
\label{eq:f0def}
\end{equation}
By construction, $f^{K \pi}_0(0)=f^{K \pi}_+(0)$. 
Since $f^{K \pi}_+(0)$ is not directly measurable, 
it is convenient to factor out it in Eq. (\ref{eq:Mkl3}) and therefore we 
introduce the normalized form factors $\bar f_{+,0}(t) \equiv f^{K \pi}_{+,0}(t) / f_{+}^{K \pi}(0)$ with 
$\bar f_+(0)=\bar f_{0} (0)=1$. 
Finally, $\Delta^{K \ell}_{\rm EM}$ represents the channel-dependent long distance electromagnetic 
corrections and $\Delta^{K \pi}_{\rm SU(2)}$ the correction induced by strong isospin breaking. 

To extract $|V_{us}|$ from the $K_{\ell 3}$ decays using Eq. (\ref{eq:Mkl3}), 
one has to measure the $K_{\ell 3}$ decay rates, to compute the phase space integrals from the 
form factor measurements and to use the theoretical estimates of $\Delta^{K \ell}_{\rm EM}$, 
$\Delta^{K \pi}_{\rm SU(2)}$ and $f_+(0) \equiv f_{+}^{K^0 \pi^-} (0)$. 
In the following, we will review the evaluation of these different ingredients. 

\subsubsection{Electromagnetic effects}
\label{sec:EMcorr}
The long distance electromagnetic corrections $\Delta^{K \ell}_{\rm EM}$ entering Eq. (\ref{eq:Mkl3}) 
were estimated only model-dependently \cite{Ginsberg:66, Bytev:2002nx, Andre:04} until recently where 
they were calculated in the framework of Chiral 
Perturbation Theory (ChPT) including photons and leptons to order 
$e^2 p^2$ \cite{Cirigliano:01,Cirigliano:2008wn}. 
To this order, both virtual and real photon corrections 
contribute. The virtual photon corrections consist in loops 
and tree level diagrams 
with an insertion of $\mathcal{O} (e^2 p^2)$ unknown low energy constants (LECs) which have to be estimated 
relying on models or lattice calculations.   
In Ref. \cite{Cirigliano:2008wn}, a fully inclusive prescription for 
real photon emission has been used as well as the more recent determinations of the LECs 
\cite{Ananthanarayan:2004qk,DescotesGenon:2005pw} based on large $N_C$ calculations. 
The results are reported in Tab.~\ref{tab:Kl3radcorr} 
and compared to previous determinations for the neutral modes
from Ref. \cite{Andre:04}. 
The errors quoted for Ref. \cite{Cirigliano:2008wn} in Tab.~\ref{tab:Kl3radcorr}
are estimates of (only partially 
known) higher order contributions. 
As can been seen, the values agree within $1 \sigma$. 
The use of ChPT has allowed to improve the previous determinations 
where the ultra violet divergences of the loops 
were regulated with a cut-off. 
\begin{table}[ht]
\setlength{\tabcolsep}{3.8pt}
\centering
\begin{tabular}{|c|c|c|c|c|}
\hline
 & $K^{0}_{e 3}$ & $K^{\pm}_{e3}$ & $K^{0}_{\mu 3}$ & $K^{\pm}_{\mu 3}$ \\
\hline
$\Delta^{K \ell}_{\rm EM}(\%)$ \cite{Cirigliano:2008wn} & 0.495 $\pm$ 0.110 & 0.050 $\pm$ 0.125 & 0.700 $\pm$ 0.110 & 0.008 $\pm$ 0.125 \\
\hline
$\Delta^{K \ell}_{\rm EM}(\%)$ \cite{Andre:04} & 0.65 $\pm$ 0.15 & - & 0.95$\pm$ 0.15 & - \\
\hline
\end{tabular}
\caption{Summary of the electromagnetic corrections
to the fully-inclusive 
$K_{\ell 3(\gamma)}$ rate~\cite{Cirigliano:2008wn} compared to the ones from the model of Ref.~\cite{Andre:04}.} 
\label{tab:Kl3radcorr}
\end{table}
The electromagnetic corrections 
to the Dalitz plot densities can also be found in Ref. \cite{Cirigliano:2008wn} 
and they can be large (up to 
$ \sim 10 \%$). Therefore, they are crucial to consider 
in the experimental extraction of the 
form factor parameters. 

\subsubsection{Isospin-breaking corrections and quark mass ratios}
\label{sect:isobreak}
In Eq. (\ref{eq:Mkl3}), $f_+^{K^0 \pi^-} (0)$ is pulled out for all decay channels, 
factorizing the isospin-breaking corrections in the $\Delta^{K \pi}_{\rm SU(2)} 
\equiv f_+^{K \pi} (0) / f_+^{K^0 \pi^-} (0) -1$ 
term. 
At leading order ($\mathcal{O}(p^2)$) in ChPT, $\Delta_{\rm SU(2)}^{K \pi}$ 
is only due to $\pi^0-\eta$ mixing and is thus proportional to 
the quark mass ratio $R$, $\Delta_{\rm SU(2)}^{K \pi }= 3/(4~R)$ \cite{Gasser:1984ux}
with $R \equiv (m_s - \widehat{m})/(m_d - m_u)$, $\widehat{m}$ being the average of the $u$ and 
$d$ quark masses. 
At NLO of the chiral expansion ($\mathcal{O}(p^4)$), 
it reads \cite{Cirigliano:2007zz}
\be
\Delta_{\rm SU(2)}^{K \pi }= \frac{3}{4} 
\frac{1}{R} \left(1 + \chi_{p^4} 
+ \frac{4}{3} 
\frac{M_K^2-M_\pi^2}{M_\eta^2-M_\pi^2} \Delta_M + \mathcal{O} (m_q^2)\right)~,
\label{eq:Rextrac} 
\ee
where $\chi_{p^4}=0.219$ \cite{Gasser:1984ux} is an $\mathcal{O}(p^4)$ ChPT correction and 
$\Delta_M$ is a correction related to the ratios of quark masses through
\beqa
\frac{M_K^2}{M_\pi^2} = \frac{m_s+\hat{m}}{m_u+m_d} \Big{(} 1 + \Delta_M + \mathcal{O} (m_q^2) \Big{)} 
\equiv \frac{Q^2}{R} \Big{(} 1 + \Delta_M + \mathcal{O} (m_q^2) \Big{)}~,
\eeqa
$Q^2 \equiv (m_s^2- \hat{m}^2) / (m_d^2- m_u^2)$ being the quark mass double ratio. 
Hence, to determine $\Delta_{\rm SU(2)}^{K \pi}$, the crucial 
inputs are the quark mass ratios $R$ and $Q^2$. One can extract them from the analysis of the 
$\eta \rightarrow 3 \pi$ decays or from the kaon mass splitting. 
A recent analysis using the latter method and including the electromagnetic effects at $\mathcal{O}(p^2 e^2)$ 
entering $\Delta_{\rm SU(2)}^{K \pi}$ obtains \cite{Kastner:2008ch}
\be
R = 33.5 \pm 4.3~,~\rm{and}~~\Delta_{\rm SU(2)}^{K \pi} = 0.029 \pm 0.004~.
\label{deltaSU2KN}
\ee
This is based on an evaluation of the electromagnetic low-energy couplings 
\cite{Ananthanarayan:2004qk} leading to a large deviation of the Dashen's limit \cite{Dashen:1969eg}.  
Note that the analyses of the $\eta \rightarrow 3 \pi$ decays \cite{Kambor:1995yc,Anisovich:1996tx,Bijnens:07} 
give results for $R$ which are higher and thus an isospin correction smaller by around $1.2 \sigma$. 
Indeed an update of Ref. \cite{Anisovich:1996tx} where $R$ was found to be $R=40.8 \pm 3.2$ 
presented in Ref. \cite{Cirigliano:2007zz} gives 
$\Delta_{\rm SU(2)}^{K \pi} = 0.0236 \pm 0.0022$. 
New analyses of this decay based on recent data \cite{Ambrosino:2008ht} are in progress 
\cite{Colangelo:2009db,Kampf} and should shed light on this issue. 
It is also possible to measure the quark mass ratios on the lattice, 
see Ref. \cite{Leutwyler:09} for a recent overview. 
For instance, taking the results from MILC collaboration \cite{Heller:09}, one gets $R=33.3 \pm 2.8$ 
in good agreement with Eq. (\ref{deltaSU2KN}).

\subsubsection{Determination of the phase space integrals and parametrization of the form factors}

The last ingredient in order to extract $|f_+(0) V_{us}|$ from Eq. (\ref{eq:Mkl3}) is 
the calculation of the phase space integrals, $I^\ell_{K}$.
To that aim, one needs to determine the normalized vector and scalar form factors, $\bar{f}_+(t)$ and $\bar{f}_0(t)$, 
entering this calculation. 
This can be achieved by a fit to the measured distributions of the $K_{\ell 3}$ decays 
assuming a parametrization for the form factors\footnote{In the Dalitz plot density formula, 
$\bar f_0$ is multiplied by a kinematic factor $(m_\ell/M_K)^2$, making it only accessible in the $K_{\mu 3}$ 
decay mode.}.
Different experimental analyses of $K_{\ell 3}$ data have been performed in the last few years, 
by KTeV \cite{KTeVe,KTeVmu}, NA48 \cite{NA48e, NA48mu}, 
and KLOE \cite{KLOEe, KLOEmu} for the neutral mode and by ISTRA+ \cite{ISTRA} for the charged mode. 

Among the different parametrizations available, one can distinguish two classes~\cite{KTeVmu}.
The class called class II in this reference contains parametrizations based on mathematical 
rigorous expansions where the slope, the curvature 
and all the higher order terms of the expansion are free parameters of the fit. 
In this class, one finds the Taylor expansion
\begin{eqnarray}
\bar{f}_{+,0}^{Tayl}(t) &=& 1 + \lambda'_{+,0} \frac{t}{M_\pi^2} + \frac{1}{2}\lambda''_{+,0} \left(\frac{t}{M_\pi^2}\right)^2 
+ \frac{1}{6}\lambda'''_{+,0} \left(\frac{t}{M_\pi^2}\right)^3 + \ldots~,
\label{Taylor}
\end{eqnarray}
where $\lambda'_{+,0}$ and $\lambda''_{+,0}$ are the slope and the curvature of the form factors respectively, 
but also the so-called z-parametrization~\cite{Hill:2006bq,capr09}.

As for parametrizations belonging to class I, they correspond to parametrizations for which by using 
physical inputs, specific relations between the slope, the curvature and all the higher order terms of 
the Taylor expansion, Eq.~(\ref{Taylor}), are imposed. 
This allows to reduce the correlations between the fit parameters since 
only one parameter is fitted for each form factor. In this class, one finds the pole parametrization 
$\bar{f}_{+,0}^{Pole}(t) = M_{V,S}^2 / (M_{V,S}^2-t)$
in which dominance of a single resonance is assumed and its mass $M_{V,S}$ is the fit parameter. 
Whereas for the vector form factor a pole parametrization with the dominance of 
the $K^*(892)$ ($M_V \sim 892$ MeV) is in 
good agreement with the data, for the scalar form factor there is not such an obvious dominance. 
One has thus to rely, at least for $\bar f_0(t)$, on a dispersive parametrization which preserves analyticity and 
unitarity. 

\paragraph{The Callan-Treiman Theorem}
Measuring the scalar form factor is also of special interest 
due to the existence of the Callan-Treiman (CT) theorem \cite{Callan:1966hu} 
which predicts the value of the scalar form factor at the so-called CT
point, $t\equiv \Delta_{K \pi}= M_K^2-M_\pi^2$,
\begin{equation}
C \equiv \bar f_0(\Delta_{K\pi})=\frac{f_K}{f_\pi}\frac{1}{f_+(0)}+ \Delta_{CT},
\label{eq:CTrel}
\end{equation}
where $f_{K, \pi}$ are the kaon and pion decay constants respectively and 
$\Delta_{CT} \sim  {\mathcal{O}} (m_{u,d}/4 \pi F_{\pi})$ is a small correction.  
ChPT at NLO in the isospin limit~\cite{Gasser:1984ux} gives\footnote{A complete two-loop calculation 
of $\Delta_{CT}$~\cite{Bijnens:2007xa}, as well as a computation 
at $\mathcal{O}(p^4, e^2p^2, (m_d-m_u))$\cite{Kastner:2008ch} 
give results consistent with this estimate.}
\begin{equation}
\label{eq:DeltaCT}
\Delta_{CT}=(-3.5\pm 8)\times 10^{-3}~,
\end{equation}
where the error is a conservative estimate of the higher order corrections~\cite{Leutwyler}. 

This correction is small enough that the right-hand side of 
Eq. (\ref{eq:CTrel}) can be determined with sufficient accuracy 
from branching ratio measurements or 
lattice QCD calculations, as will be discussed in section \ref{sec:CTtest}, 
to compare with $C$ measured in $K_{\mu 3}$ decays. 
Thus apart from the determination of $|V_{us}|$ which is used to test the 
unitarity of the CKM matrix within the SM, a measurement of $C$ 
provides an other interesting test of the SM namely a test 
of the couplings of the light quarks to the W-boson. 
It can also give constraints on the presence of charged Higgs 
or charged right-handed current couplings, see section \ref{sec:CTtest}.

Another interest in the experimental determination of the 
shape of $\bar f_0$ is the possibility of determining some LECs appearing in 
ChPT \cite{Bernard:2007tk}.


\paragraph{Dispersive parametrization for the form factors}
Motivated by the existence of the CT theorem a dispersive parametrization for the scalar form factor 
has been proposed in Ref.~\cite{Bernard:2006gy}. 
Two subtractions are performed, one at $t=0$ where by definition $\bar f_{0}(0)=1$, 
and the other one at the CT point. 
%
Assuming that the scalar form factor has no zero, one can write 
\begin{equation}
\bar f_0^{Disp}(t)=\exp\Bigl{[}\frac{t}{\Delta_{K\pi}}(\mathrm{ln}C- G(t))\Bigr{]}~, 
\rm{with}~~
G(t)=\frac{\Delta_{K\pi}(\Delta_{K\pi}-t)}{\pi} \int_{(M_K+M_\pi)^2}^{\infty}
\frac{ds}{s}
\frac{\phi_0(s)}
{(s-\Delta_{K\pi})(s-t-i\epsilon)}~.
\label{eq:Dispf}
\end{equation}
In this case the only free parameter to be determined from a fit to the data is ln$C$, 
the logarithm of $C$, Eq. (\ref{eq:CTrel}). 
$\phi_0(s)$ represents
the phase of the form factor. According to Watson's theorem
\cite{Watson:1952ji}, this phase can be identified in the elastic region with the S-wave, $I=1/2$ $K\pi$ scattering phase. 
The fact that two subtractions have been made in writing Eq.~(\ref{eq:Dispf}) allows
to minimize the contributions from the unknown high-energy
phase (taken to be $\pi \pm \pi$) in the dispersive integral. 
The resulting function $G(t)$, Eq.~(\ref{eq:Dispf}),
does not exceed 20\% of the expected value of ln$C$ limiting the
theoretical uncertainties which represent at most 10\% of the value of
$G(t)$ \cite{Bernard:2006gy,Bernard:2009zm}.

A dispersive representation for the vector form factor has been built in a 
similar way \cite{Bernard:2009zm}. 
Since there is no analog of the CT theorem, in this case, the two subtractions are performed at $t=0$. 
Assuming that the vector form factor has no zero, one gets
\begin{equation}
\bar f_+^{Disp}(t)=\exp\Bigl{[}\frac{t}{M_\pi^2}\left(\Lambda_+ + H(t)\right)\Bigr{]}~,
\label{eq:Dispfp}
\rm{with}~~
H(t)=\frac{M_\pi^2t}{\pi} \int_{(M_K+M_\pi)^2}^{\infty}
\frac{ds}{s^2}
\frac{\phi_+ (s)}
{(s-t-i\epsilon)}~. 
\end{equation}
$\Lambda_+~\equiv M_\pi^2 d \bar f_+(t)/dt|_{t=0}$ is the fit parameter and
$\phi_+(s)$ the phase of the vector form factor. Here, in the elastic region, $\phi_+(t)$ equals 
the P-wave, $I=1/2$ K$\pi$ scattering phase according to Watson's theorem \cite{Watson:1952ji}.  
Similarly to what happens for $G$, the two subtractions minimize the contribution coming from the 
unknown high energy phase resulting in a relatively small uncertainty on $H(t)$. 
Since the dispersive integral $H(t)$ represents at most 20\% of the expected value of $\Lambda_+$, 
the latter can then be determined with a high precision knowing $H(t)$ much less precisely. 
For more details on the dispersive representation and a 
detailed discussion of the different sources of theoretical uncertainties 
entering it via the functions $G$ and $H$,
see Ref. \cite{Bernard:2009zm}.
\\
\\
Using a class II parametrization  for the form factors in a fit to $K_{\ell 3}$ decay distribution, 
only two parameters ($\lambda_+'$ and $\lambda''_+ $ for a Taylor expansion, 
Eq.~(\ref{Taylor}))
%
%
can be determined for $\bar f_+(t)$ and only one  parameter
($\lambda'_0$ for a Taylor expansion)
for $\bar f_0(t)$. 
Moreover these parameters are strongly correlated. 
It has also been shown in Ref. \cite{Bernard:2006gy} that in order to describe 
the form factor shapes accurately in the physical region, 
one has to go at least up to the second order in the Taylor expansion. 
Neglecting the curvature in the parametrization of $\bar f_0(t)$  
generates a bias in the extraction of $\lambda_0'$ which is then overestimated \cite{Bernard:2006gy}.  
Hence, using a class II parametrization  for $\bar f_0(t)$ doesn't allow 
for extrapolating it from the physical region ($m_{\ell}^2 < t < (M_K-M_\pi)^2$) 
up to the CT point with a reliable precision.  
To measure the form factor shapes from $K_{\ell3}$ decays with the precision 
demanded in the extraction of $|V_{us}|$, it is thus preferable to use 
a parametrization in class I and thus for all the reasons given before the dispersive parametrization 
of Eqs. (\ref{eq:Dispf}) and (\ref{eq:Dispfp}) for the form factors. 

\begin{figure}[h]
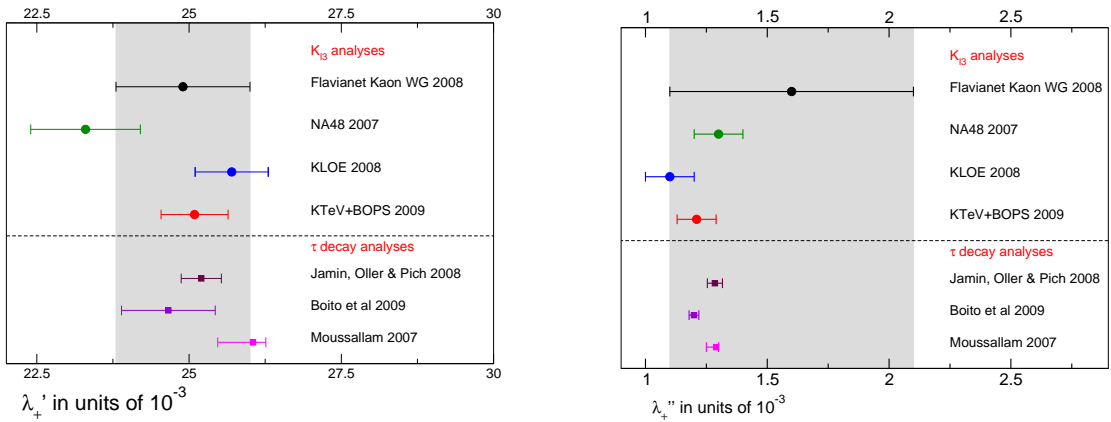

\includegraphics[scale=0.30]{fig1MT.eps}
\includegraphics[scale=0.30]{fig2.eps}
\caption{Results on $\lambda'_+$ (left) and $\lambda''_+$ (right) from the dispersive 
analyses of the $K_{\ell 3}$ decays \cite{NA48mu, KLOEmu, KTeVmu} compared 
to the average of quadratic/linear 
analyses from the Flavianet Kaon Working Group (WG) \cite{KWG} and to the results from $\tau \rightarrow K \pi \nu_\tau$ 
analyses \cite{Jamin:2008qg, Boito:2008fq,Bachir08}.}
\label{fig:lambdasV}
\end{figure}

\begin{figure}[t]
\begin{minipage}{9.cm}
\vspace{-5.cm}
\hspace{-0.4cm}
\begin{tabular}{ c c c c }
\hline
\hline
Experiment & Channel & ln$C$ & $r$ \\
\hline
KTeV \cite{KTeVmu} & K$_{L \mu3}$+K$_{L e 3}$ & 0.192(12) & 0.978(14) \\
\hline
KLOE \cite{KLOEmu} & K$_{L \mu3}$+K$_{L e 3}$ & 0.204(25) & 0.990(26) \\
\hline
NA48 \cite{NA48mu} & K$_{L \mu3}$ & 0.144(14) & 0.932(15) \\
\hline
SM & & 0.2141(73) & 1 \\
\hline
\hline
\end{tabular}
\end{minipage}
\hfill
\hspace{-0.4cm}
\includegraphics[scale=0.30]{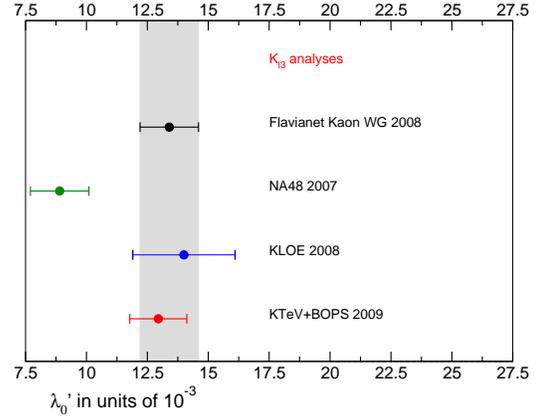}
\caption{Left: Table, Results on ln$C$ from the dispersive 
analyses of the $K_{\ell 3}$ decays \cite{NA48mu, KLOEmu, KTeVmu} compared to the SM expectation 
and extraction of $r$, Eq. (3.6). 
The SM expectation for ln$C$ is obtained from Eq. (3.4) using $(f_K/(f_\pi f_+(0)))|_{SM}$, Eq. (3.5), and $\Delta_{CT}$, 
Eq. (2.9), as input values. 
Right: Results on $\lambda'_0$ from the dispersive 
analyses of the $K_{\ell 3}$ decays \cite{NA48mu, KLOEmu, KTeVmu} compared 
to the average of the quadratic/linear 
analyses from the Flavianet Kaon WG \cite{KWG}.}
\label{fig:lambda0}
\end{figure}

\paragraph{Results of the dispersive analyses}
A dispersive analysis has now been performed by NA48 \cite{NA48mu}, KLOE \cite{KLOEmu} 
and KTEV \cite{KTeVmu} for their $K_{L \ell 3}$ decays. 
The results are presented in Fig.\ref{fig:lambdasV} calculating 
from Eq. (\ref{eq:Dispfp}) the slope and curvature of 
the vector form factor and in Fig.\ref{fig:lambda0} 
for the slope of the scalar one calculated from Eq. (\ref{eq:Dispf}). 
For comparison, the results coming from an average of the fits 
using a quadratic Taylor expansion for $\bar{f}_+(t)$ and a linear one for $\bar{f}_0(t)$ 
\vspace{0.3cm}
(quadratic/linear analysis)
from the Flavianet Kaon Working Group (WG) \cite{KWG} are also represented.
It can be seen that for the vector form factor 
the results of the dispersive analyses of KLOE, KTeV and NA48 agree reasonably well for $\lambda'_+$ and 
very well for $\lambda''_+$.  
The dispersive analyses improve very much on the precision of the measured slope 
and curvature compared to the previous 
quadratic/linear ones; especially the result 
on the curvature 
is improved by a factor 4 compared to the average of the quadratic/linear fits. 
The vector form factor can also be determined from studies of $\tau \rightarrow K \pi \nu_\tau$ decays using the 
recent data from Belle \cite{Belle:07} and BaBar \cite{Babar:07}. 
The results using Belle data are shown in Fig.\ref{fig:lambdasV}. 
To describe the $\tau$ decays, one needs to know the phase of the vector form factor 
at higher energy than for the $K_{\ell 3}$ decay fits. To determine it in the inelastic region, 
one has thus to rely on models. For instance, in Refs. \cite{Jamin:2008qg, Boito:2008fq} the inelastic effects have been neglected and 
two resonances have been included and in Ref. \cite{Bachir08} a coupled-channel analysis has been performed. 
As can be seen from Fig.\ref{fig:lambdasV}, there is a very good agreement with the $K_{\ell 3}$ analyses 
for both determination of the slope and curvature with a much smaller uncertainty on the determination of 
the curvature. 
A combination of the two analyses, as presented in Ref. \cite{Boito:2009pv}, 
can be very interesting to perform in order to have a better determination of the vector form factor parameters and to test 
the correlation between slope and curvature imposed by the dispersion relation used in the $K_{\ell 3}$ analyses. 

As for the determination of the scalar form factor, 
the results on its slope shown in figure 2 (right) from KLOE and KTeV are in agreement whereas 
the one from NA48 differs from the others by more than 2 $\sigma$. $\lambda_0''$ is found to be small, of
order $6 \cdot 10^{-4}$ but it can't be neglected to reach a high degree of precision. 
In principle, a measurement of the scalar form factor is also possible  
from the lowest energy part of the $\tau \rightarrow K \pi \nu_\tau$ decay distribution where it dominates. 
But, at present, the precision at threshold is not high enough to allow for a competitive determination of 
the scalar form factor parameters from this decay. A measurement of $\bar f_0(t)$ 
from $\tau \rightarrow K \pi \nu_\tau$ decays with a good precision 
will certainly be possible in the near future thanks to new measurements, 
see for example Ref. \cite{Paramesvaran:2009ec}. 
Moreover $K^+$ measurements from NA48 and KLOE are underway 
which might help to solve the puzzle on the scalar form factor measurement. 

\subsubsection{Determination of $f_+(0)$}
\begin{figure}[t]
\hspace{-1.cm}
\includegraphics[width=0.6\textwidth]{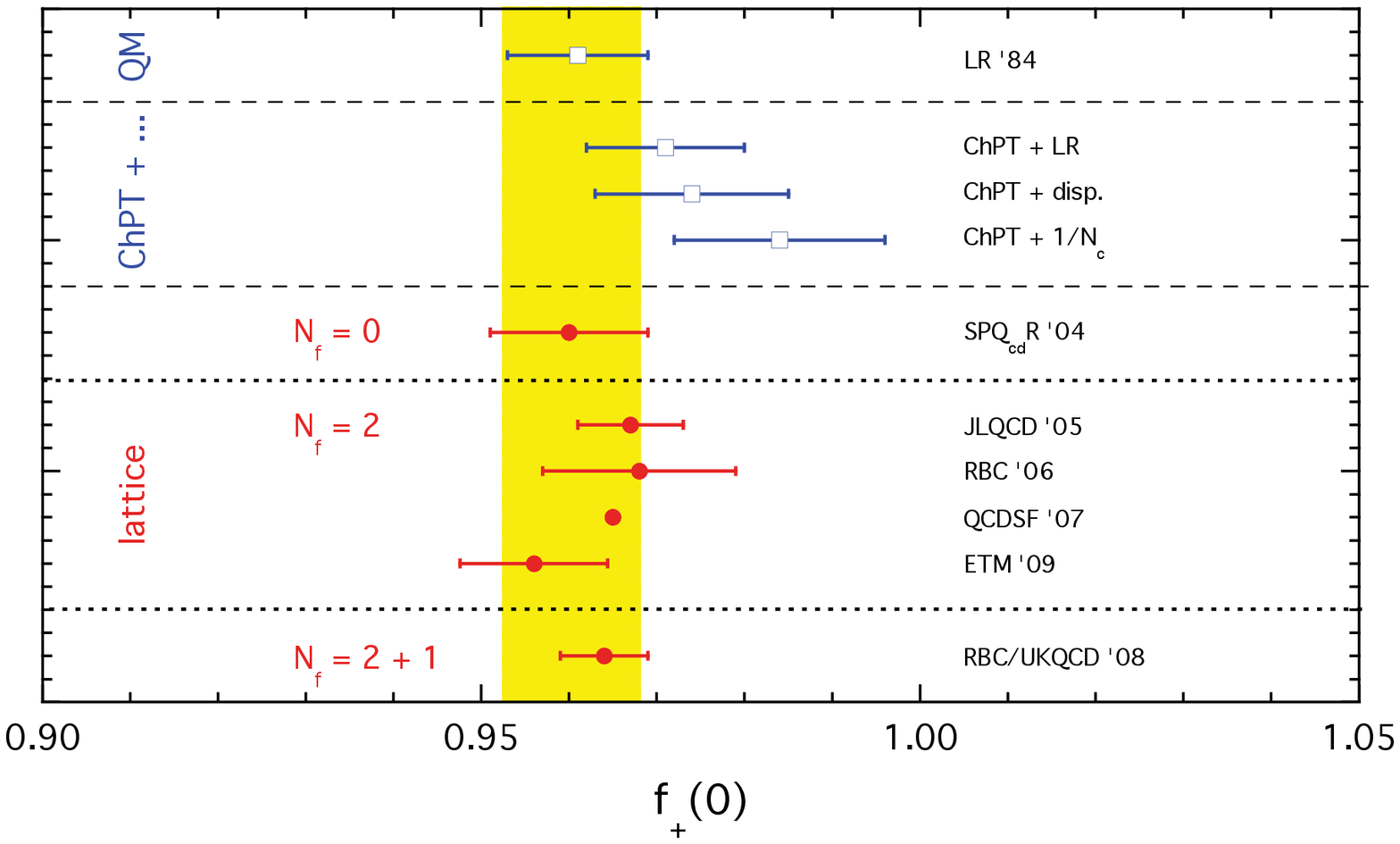}
\hspace{-1.cm}
\includegraphics[width=0.6\textwidth]{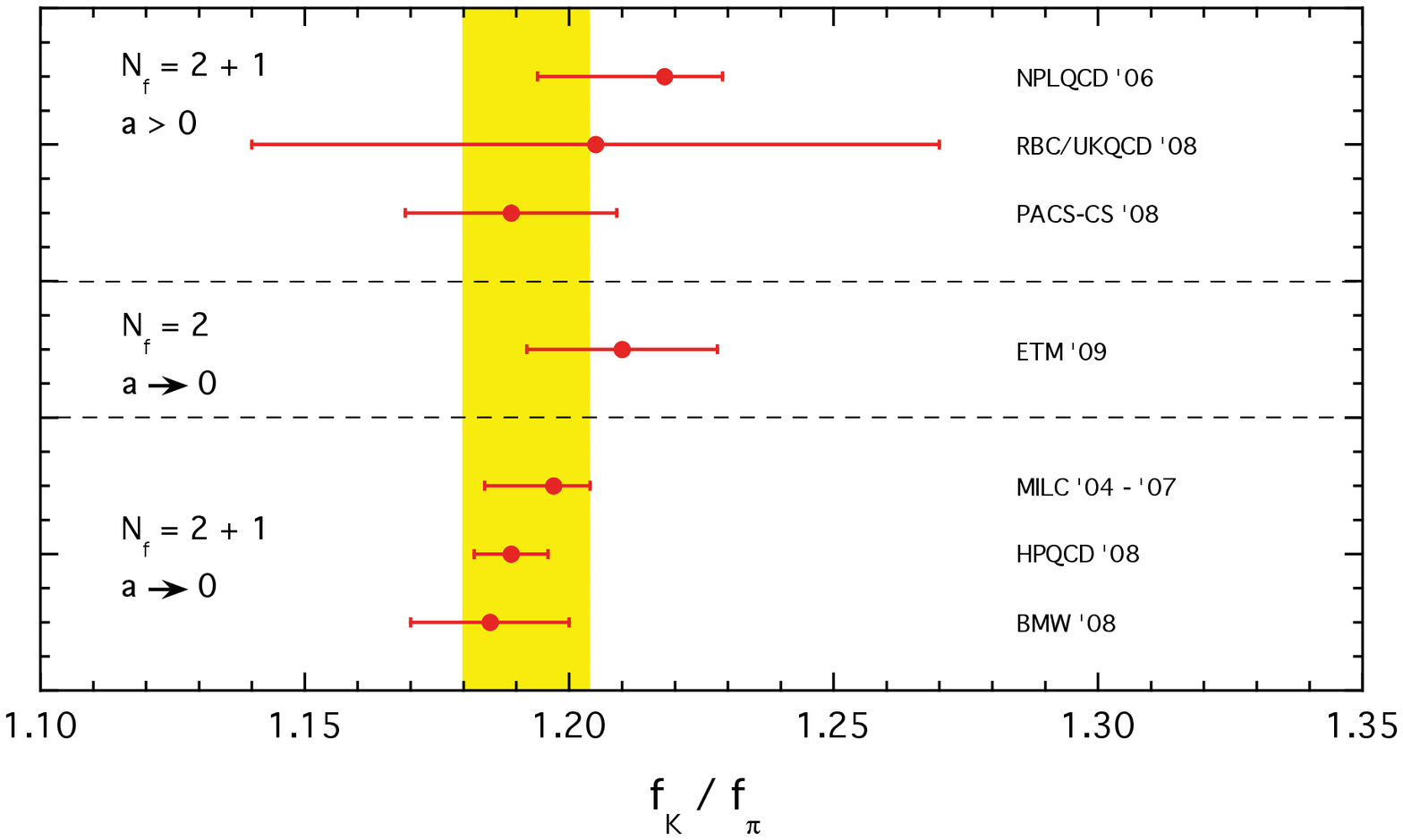}
\vspace{-8.cm} 
\caption{Left: Present determinations 
of $f_+(0)\equiv f_+^{K^0\pi^-}(0)$ in blue from analytical or semi-analytical 
approaches~\cite{Leutwyler:1984je,Bijnens:2003uy,Jamin:2004re,Cirigliano:2005xn,Kastner:2008ch} 
compared to the ones in red from lattice
QCD \cite{Becirevic:2004ya,Tsutsui:2005cj,Brommel:2007wn,Lubicz:2009ht,Boyle:2007qe}. 
Right: $f_K/f_\pi$ from lattice QCD calculations \cite{Blossier:2009bx,
Aubin:2004fs,Beane:2006kx,Follana:2007uv,Allton:2008pn,Aoki:2008sm,Aubin:2008ie,Durr2008}. 
The yellow bands correspond to average values given in \cite{CKMbook,FLAG} and 
$N_{f}$ denotes the number of dynamical quarks. 
Figures courtesy of S. Simula.}
\label{fig:f0}
\end{figure}

Using the theoretical inputs described before and the most recent experimental data \cite{KWG}, from Eq. (\ref{eq:Mkl3}), 
one gets as an average on the neutral and charged modes \cite{KWGT}
\be
|f_+(0) V_{us}|= 0.2163(5)~.
\ee
To determine $|V_{us}|$, one has to calculate the hadronic key quantity $f_+(0)$.  
In the chiral limit and, more generally, in the $SU(3)$ limit
($m_u=m_d=m_s$) the conservation of the vector current (CVC) 
implies $f_+(0)$=1. Expanding around the chiral limit in powers 
of light quark masses we can write
\be
\label{eq:chpt4}
f_+(0)= 1 + f_2 + f_4 + \ldots,
\ee
where $f_n = \mathcal{O} \left( m_{u,d,s}^n / (4 \pi f_\pi)^n \right)$,  
$f_2$ and $f_4$ being the NLO and 
NNLO corrections in ChPT.  The Ademollo-Gatto theorem \cite{Ademollo} implies that
$[f_+(0)-1]$ is at least of second order in the breaking of $SU(3)$ so that 
$f_2$ is free of any LECs. 
It can therefore be computed with high accuracy: 
$f_2=-0.0227$~\cite{Gasser:1984ux,Leutwyler:1984je}. 
The difficulty is then in the calculation of the quantity $\Delta f$ 
\be
\Delta f \equiv f_4 + f_6 + \ldots = f_+(0) - (1 + f_2)~,
\ee
which depends on some LECs. 
The original estimate based on quark model from Leutwyler and Roos (LR) \cite{Leutwyler:1984je} 
gives $\Delta f = -0.016(8)$ and $f_+(0) = 0.961 (8)$. 
Then more recently some analytical calculations have been performed to evaluate the 
NNLO term $f_4$ writing it as  
\be 
\renewcommand{\arraystretch}{0.5} 
f_4 = \Delta(\mu) + f_4\vert^{\rm loc}(\mu)\,, 
\label{eq:f4ch}
\ee 
where $\Delta(\mu)$ is the loop contribution, $\mu$ being the renormalization scale 
which has been computed in Ref.~\cite{Bijnens:2003uy}, and $f_4\vert^{\rm loc}(\mu)$ is the 
$\mathcal{O}(p^6)$ local contribution which contains unknown $\mathcal{O}(p^4)$ and 
$\mathcal{O}(p^6)$ LECs. To estimate the latter term various models have been used 
namely a quark model \cite{Bijnens:2003uy}, 
dispersion relations \cite{Jamin:2004re} or $1/N_C$ estimates \cite{Cirigliano:2005xn}.
These calculations 
allow for a better control of the systematic uncertainties. 
The values obtained are summarized in Fig.\ref{fig:f0} (left). They are significantly 
larger than the LR estimate leading to smaller $SU(3)$ breaking effects on $f_+(0)$. 




Note that in principle the NNLO term $f_4$ may be obtained from the measurement of the slope and the curvature 
of the scalar form factor $f_0(t)$ \cite{Bijnens:2003uy,Bernard:2007tk}. 
The present level of precision coming from the $K_{L \mu3}$ 
measurements as well as the knowledge of the $\mathcal{O}(p^4)$ LECs entering the calculation limit the 
precision of the extraction. 
The procedure of averaging the scalar form factor parameters within the Flavianet kaon working group 
as well as the update of the determination of the $\mathcal{O}(p^4)$ LECs \cite{Bijnens:2009hy} 
should improve on the precision of this determination in the near future.  

As it was discussed at this conference, lattice calculations 
give very precise results for $f_+(0)$, up to the level of $0.5 \%$ as shown in Fig.\ref{fig:f0} (left). They are in 
agreement with the result from the pioneering work of LR \cite{Leutwyler:1984je} but give 
larger $SU(3)$ breaking compared to the recent analytical estimates. This is essentially due to the large positive 
two loop contribution \cite{Bijnens:2003uy}. 
Note that only one calculation for $f_+(0)$ in $N_f = 2+1$ flavours exists. Using this result \cite{Boyle:2007qe} leads to \cite{KWGT},
$|V_{us}|=0.2243(12)$,
which is at the moment the most precise extraction of $|V_{us}|$. 

\subsection{$K_{\ell 2}/ \pi_{\ell 2}$ ratio of decay rates}
Another possibility to extract $|V_{us}|$ is to use the photon inclusive $K_{\ell2}/\pi_{\ell2}$ 
ratio of decay rates written as \cite{Marciano:1993sh}
\begin{eqnarray}
\frac{\Gamma_{K^{\pm}_{\ell 2(\gamma)}}}{\Gamma_{\pi^{\pm}_{\ell 2(\gamma)}}} =
\left|\frac{V_{us}}{V_{ud}} \right|^2\frac{f^2_K M_K}{f^2_\pi M_\pi}
\left(\frac{1-m^2_\ell/M_K^2}{1-m^2_\ell/M_\pi^2}\right)^2~\left(1+\delta_{\rm em}\right), 
\label{eq:Mkl2}
\end{eqnarray}
where $\delta_{\rm em}$ stands for the long-distance electromagnetic (EM) corrections 
and depends on the hadronic structure and the particle masses as well as on the electromagnetic 
radiative corrections. For more details on its calculation, see Ref. \cite{Marciano:1993sh}. 
By taking the ratio, $\Gamma_{K^{\pm}_{\ell 2(\gamma)}}/\Gamma_{\pi^{\pm}_{\ell 2(\gamma)}}$, Eq. (\ref{eq:Mkl2}), 
all the channel-independent corrections cancel allowing 
to estimate the EM corrections with a good precision, 
$\delta_{\rm em}=-0.0070(35)$ \cite{Marciano:2004uf}. 
Hence from the experimental measurements of the ratio of decay rates 
$\Gamma_{K^{\pm}_{\ell 2(\gamma)}}/\Gamma_{\pi^{\pm}_{\ell 2(\gamma)}}$, 
one can extract \cite{CKMbook}
\be
\frac{f^2_K}{f^2_\pi} \left|\frac{V_{us}}{V_{ud}} \right|^2 = 0.2758 \pm 0.0007~.
\ee  
To access $|V_{us}|/|V_{ud}|$, one needs to know 
the ratio of the kaon versus pion decay constants $f_K/f_\pi$. 
The analytic evaluation of this ratio within ChPT 
depends already at order $\mathcal{O} (p^4)$ on the determination of some LECs 
and thus bring additional uncertainties. 
This is why the most precise evaluations come from lattice QCD calculations. 

The status of the 
lattice results for $f_K/f_\pi$ is summarized in Fig.\ref{fig:f0}~(right). 
The agreement between the different results is remarkable and the 
present overall accuracy is very good being of about $1\%$. 
For discussions and details on the lattice set-ups and results, 
we refer to the forthcoming review by the Flavianet Lattice Averaging Group (FLAG) 
presented at this conference by G. Colangelo \cite{FLAG} and also to the talk by P.~Boyle \cite{Boyle:09}. 
As presented in Ref. \cite{Palutan}, taking for illustrative purpose the HPQCD-UKQCD result \cite{Follana:2007uv}, 
one 
gets $|V_{us}|/|V_{ud}|=0.2319(15)$.

\section{Applications: Stringent tests of the Standard Model}
\subsection{Determination of $|V_{us}|$ and the CKM unitarity test}
Combining the result on $|V_{us}|$ from $K_{\ell 3}$ decays, with the one 
on $|V_{us}|/|V_{ud}|$ from the $K_{\ell 2}/\pi_{\ell 2}$ ratio of decay rates 
and the one on $|V_{ud}|$ from superallowed $\beta$ decays \cite{Hardy:2008gy} in a global fit allows to improve 
on the precision of the determination of $|V_{us}|$ and to perform 
a very precise test of the CKM unitarity (test of unitarity of the first row of the CKM matrix) which leads to stringent constraints 
on new physics scenarios \cite{KWG, Marciano:2007zz, Cirigliano:2009wk}, 
see the talk by M. Palutan \cite{Palutan}. 
\subsection{Test of lepton universality and extraction of the quark mass ratio using $K_{\ell 3}$ decays}
The precision reached in the measurements of the $K_{\ell 3}$ decays \cite{KWG} 
and in the determination of the EM corrections, see section 
\ref{sec:EMcorr}, allows for putting very interesting constraints on lepton universality and also 
for a precise extraction of the quark mass ratios, $Q$ and $R$. 
Concerning the test of lepton universality, using Eq. (\ref{eq:Mkl3}), one can determine the ratio 
\be
r_{\mu / e} = \frac{G_\mu^2}{G_e^2} = \frac{\Gamma_{K_{\mu 3}}}{\Gamma_{K_{e 3}}} 
\frac{I_{K}^{e}}{I_{K}^{\mu}} (1+2 \Delta_{\rm EM}^{K e} - 2 \Delta_{\rm EM}^{K \mu})~.
\ee 
Averaging charged and neutral modes, one gets $r_{\mu / e} = 1.002 \pm 0.005$ \cite{KWGT} 
consistent with lepton universality. 
With an accuracy of $0.5 \%$, this determination becomes by now competitive 
with other determinations such as the one coming from $\tau$ decays, see Ref. \cite{Palutan}. 

In addition, a phenomenological determination of $\Delta_{\rm SU(2)}^{K \pi}$, see 
section \ref{sect:isobreak}, is by now possible comparing the results on charged and neutral decay widths, Eq. (\ref{eq:Mkl3}). 
One obtains \cite{KWG, Palutan}
\be
\Delta_{\rm SU(2)}^{K \pi} = \frac{\Gamma_{K^+_{\ell 3}}}{\Gamma_{K^0_{\ell 3}}} \frac{I_{K^0}^\ell}{I_{K^+}^\ell} 
\left( \frac{M_{K^0}}{M_{K^+}} \right)^5 -\frac{1}{2} - [\Delta^{K^+ \ell}_{\rm EM}- \Delta^{K^0 \ell}_{\rm EM}] =
 2.7(4) \%
\ee 
in good agreement with the value of Eq. (\ref{deltaSU2KN}), favouring a small value for R
\be
R=33.2 \pm 4.7~.
\label{eq:Rexp}
\ee
\subsection{Test of the SM electroweak couplings via the CT theorem with $K_{\ell 3}$ and $K_{\ell 2}$ decays}
\label{sec:CTtest}
Among the possible tests of the Standard Model and the new physics scenarios using $K_{\ell 3}$ and $K_{\ell 2}$ decays, 
the Callan-Treiman theorem, Eq. (\ref{eq:CTrel}), offers a very stringent test of the SM electroweak (EW) couplings 
due to the small size of $\Delta_{CT}$, Eq. (\ref{eq:DeltaCT}). 
The test consists of comparing the value of $f_K / (f_\pi~f_+(0))$ deduced from the measurement of C from the $K_{\mu 3}$ 
dispersive analyses using the Callan-Treiman theorem rewritten as 
\be
\frac{f_K}{f_\pi f_+(0)}=C-\Delta_{CT}~,
\label{eq:C}
\ee
to the value of $f_K / (f_\pi~f_+(0))|_{SM}$ determined by assuming the SM EW couplings and using the 
experimental measurements of $\Gamma_{K^+_{\ell 2 (\gamma)}}/\Gamma_{\pi^+_{\ell 2 (\gamma)}}$, 
Eq. (\ref{eq:Mkl2}) \cite{CKMbook}, of $\Gamma_{K_{L e3 (\gamma)}}$, Eq. (\ref{eq:Mkl3}), (from the electronic neutral mode) 
\cite{KWGT} and the value of $|V_{ud}|$ \cite{Hardy:2008gy}. One 
gets 
\be
\frac{f_K}{f_\pi f_+(0)}|_{SM} = a~\left(\frac{\Gamma_{K^{\pm}_{\ell 2(\gamma)}}}{\Gamma_{\pi^{\pm}_{\ell 2(\gamma)}}}~
\frac{1}{\Gamma_{K_{Le3 (\gamma)}}} \right)^{1/2}~|V_{ud}| = 1.2422 \pm 0.0043~,  
\label{eq:BexpSM}
\ee
with $a$, a constant determined from the theoretical inputs, see Eqs. (\ref{eq:Mkl3}) and (\ref{eq:Mkl2}). 
If the two values differ that means if 
\be
r \equiv (C-\Delta_{CT}) \cdot \frac{f_\pi f_+(0)}{f_K}|_{SM} 
\label{eq:r}
\ee
is different from one, that would indicate the presence of new physics 
such as modification of EW couplings of quarks due to exchanges 
of new particles close to the TeV scale. This can 
be, for instance, due to a direct coupling of right-handed quarks 
to the W-boson which appears at NLO of an Higgsless low-energy effective theory \cite{Bernard:2006gy,Bernard:2007cf} 
or to couplings of quarks to a charged Higgs. Hence the CT 
theorem can be used to constrain the masses and the couplings of 
the new particles appearing in such scenarios\footnote{Note that in charged Higgs scenarios, the presence of scalar couplings 
affect the ratio of decay rates $\Gamma_{K^+_{\ell 2 (\gamma)}}/\Gamma_{\pi^+_{\ell 2 (\gamma)}}$ 
as well as the extraction of ln$C$ from the $K_{\mu 3}$ decays contrary to the scenario with charged right-handed currents 
where the measurement of ln$C$ isn't affected by the new couplings, see Ref. \cite{KWG}. One has thus to put constraints 
on the charged Higgs effects by comparing the direct 
measurement of $C$ with the ratio $\left(f_K/ (f_\pi f_+(0))\right)|_{SM}$ calculated on the lattice.}, 
see e.g. Refs. \cite{Bernard:2006gy, KWG, Palutan, Deschamps09} for further discussions. 
The results from the different analyses are reported in the table 
Fig.\ref{fig:lambda0} (left) for ln$C$, the logarithm of 
$C$, Eq. (\ref{eq:C}) and for $r$, Eq. (\ref{eq:r}). The result of NA48 doesn't agree with the ones from 
KTeV and KLOE which are consistent with the SM expectation. 
It disagrees with the SM at $4.5 \sigma$ 
indicating, if confirmed, presence of physics beyond the SM. An effect at the level of several percents would be, 
however, difficult to accommodate in 
a charged Higgs scenario where the effects for natural values of the parameters are expected at the per mile level, 
whereas in the Higgless case, that would indicate an inverted 
hierarchy for the right-handed mixing matrix $V_R$. As mentioned already, some measurements 
especially for the charged modes are underway to try to solve this puzzle. 

\section{Lepton universality test via $R_K$}
Let me briefly mention a particularly sensitive probe of new physics given by the universality ratio 
\be
R_K^{e/\mu}= \frac {\Gamma_{K_{e2(\gamma)}}} {\Gamma_{K_{\mu 2(\gamma)}}}~.
\ee
This quantity is helicity suppressed in the SM as a consequence of the $V-A$ structure 
of the charged currents and thus is very small. Moreover it has been calculated with a very high level of accuracy 
recently \cite{Cirigliano:2007xi} in a first systematic calculation to $e^2 p^2$ leading to 
\be
R_K^{e/\mu} = (2.477 \pm 0.001) \times 10^{-5}~.
\label{RKSM}
\ee
The uncertainty is at the level of 0.04 $\%$ mainly due to the fact that to a first approximation 
many contributions such as the 
hadronic ones cancel out by taking the ratio. The hadronic structure dependence only appears through EW corrections. 
So one has only to evaluate the diagrams with photons connected 
to lepton lines. The relevant counterterms are estimated by a matching with large $N_C$ QCD. In this 
work, the real photon corrections have been included and the leading logarithms have been resumed. The previous 
calculation of Ref. \cite{Finkemeier:1995gi} has been improved. However a discrepancy at $5 \sigma$ exists between the two results 
but can be traced back to inconsistencies in the previous analysis. 

Measuring this ratio is very interesting since it has been shown that in realistic supersymmetric frameworks 
with new sources of lepton flavour mixings, deviations from the SM can reach the level of $1 \%$ \cite{Masiero:2005wr}. 
Since the conference Kaon'07 there have been very impressive experimental improvements in 
the measurement of this quantity as presented at this conference \cite{Collaboration:2009cya, NA62}. 
The world average from March 2009 \cite{Spadaro:2009uf} is 
\be
R_K^{e/\mu} = (2.468 \pm 0.025) \times 10^{-5}~,
\ee
confirming the SM expectation, Eq. (\ref{RKSM}). 
There is still room for improvements on the experimental side which 
could be achieved thanks to the NA62 experiment \cite{NA62}. 

\section{Conclusion} 
\vspace{-0.2cm}
The charged current analyses using $K_{\ell 3}$ and $K_{\ell 2}$ data have entered an area of very high precision. 
On the theoretical side, a lot of work has been performed within the last years to improve on the precision of 
the determination of the electromagnetic and isospin breaking corrections. Moreover dedicated dispersive parametrizations to analyse 
the $K_{\ell 3}$ form factors with a high level of accuracy have been built. 
A lot of progress has also been done 
experimentally, see Ref. \cite{Palutan}. This allows for very precise tests of the SM such as the CKM unitarity test or lepton 
flavour violation ones and gives interesting constraints on new physics scenarios. 
One should notice that in order to improve on the SM tests, especially on the one using the CT theorem, one would need to improve on the 
scalar form factor measurements and in particular on the charged channel where only one recent measurement 
has been performed \cite{ISTRA}. This would also allow for a better precision on the extraction of the quark mass ratio $R$.
On the theoretical side, the determination 
of $f_+(0)$ should be improved. Indeed, the lattice determinations don't agree well with the analytical results and there 
is only one $N_f = 2+1$ flavours result. In this respect, the recent progress in lattice calculations are promising. 

\paragraph{Acknowledgments} 
I would like to thank the organizers for their invitation and for this very pleasant and interesting conference. 
This work has been supported in part by the EU contract MRTN-CT-2006-035482 ("Flavianet").

\end{document}